\documentclass[a4paper,11pt]{article}
\usepackage{jheppub} % for details on the use of the package, please
\usepackage{bbold}
\newcommand{\be}{\begin{equation}}
\newcommand{\ee}{\end{equation}}
\newcommand{\ea}{\end{eqnarray}}
\newcommand{\ba}{\begin{eqnarray}}
\def\I{\mathbb{1}}

\newcommand{\wt}{\widetilde}

\title{\boldmath An alternative way to explain how non-commutativity arises in the bosonic string theory}

\author{M. A. De Andrade}
 \author{and C. Neves}
\affiliation{Departamento de Matem\'{a}tica, F\'{\i}sica e Computa\c{c}\~{a}o, Faculdade de Tecnologia, \\ Universidade do Estado do Rio de Janeiro,\\
Rodovia Presidente Dutra, Km 298, P\'{o}lo Industrial, CEP 27537-000, Resende-RJ, Brazil.}

\emailAdd{$\ast$deandrade.marco@gmail.com}
\emailAdd{$\dagger$clifford\_dmfc\_fat@outlook.com}

\abstract{
In this work we will investigate how the non-commutativity arises into the string theory, \textit{i.e.}, how the bosonic string theory attaches to a D3-brane in the presence of magnetic fields. In order to accomplish the proposal, we departure from the commutative two-dimensional harmonic oscillator, which after the application of the general Bopp's shifts Matrix Method, the non-commutative version of the two-dimensional harmonic oscillator is obtained. After that, this non-commutative harmonic oscillator will be mapped into the bosonic string theory in the light cone frame, which it now appears as a bosonic string theory attached to a D3-brane.
}


\begin{document}

\maketitle
\flushbottom

\section{Introduction}
\label{sec:intro}
A. Connes, in Ref.\cite{conne}, suggested that the operator algebra is like a framework for physics and, in this context, he investigated the Yang-Mills theory on a non-commutative torus. In \cite{CDS}, the authors proved, through toroidal compactification in the limit of small volume, that the NC Yang-Mills theory arises in string theory in a definite limit, which it is natural in the context of matrix model of M-theory\cite{banks1,LS,banks2,banks3,BS}. Just a week after \cite{CDS}, another paper \cite{jabbari} mention the possibility of direct noncommutativity in string theory. After that, some papers\cite{jabbari1,DH,jabbari2,ardalan} explore the connection between string theory and the Matrix theory.

 However, the physics of non-compact $\mathbb{R}^4$ is the opposite of small volume limit and, also, how to explain the strings propagation on the torus when the volume of the torus is taken to zero? In order to restore the usefulness of NC Yang-Mills theory, it was suggested that the non-commutativity of string theory is due to the presence of a $B_{\mu\nu}$-field. In \cite{jabbari3}, the author investigate the scattering of massless open strings attached to a D2-brane living in the B field background and show that corresponding scattering up to the order of $\alpha^{\prime 2}$ is exactly given by the gauge theory on non-commutative background, which it is characterized by the Moyal bracket. In \cite{schomerus} it was observed that D-branes in a presence of a non-vanishing $B_{\mu\nu}$-field leads to a Moyal deformation of the algebra of functions on the classical world-volume, which coincides with the Kontsevich's quantization formula\cite{kont}. 

In \cite{jabbari4,jabbari5} different aspects of non-commutative Yang-Mills theory within string theory and field theory setups were studied and in \cite{SW} the quantization of open strings ending on a D-branes in the presence of a $B_{\mu\nu}$-field was reexamined and also quoted that a sigma model, with a specific boundary interaction and gauge fixing terms, it is a special case of the deformed quantization theory used by Kontsevich\cite{kont}, which was well elucidated in \cite{CF}. We purchase this idea and, in this paper, we will obtain the open bosonic string attached to a D3-brane in a presence of a $B_{\mu\nu}$-field from the non-commutative version of two-dimensional harmonic oscillator, which it is obtained applying the general Bopp's shift matrix method \cite{MACN} (this result could be obtained using the non-commutative symplectic induction formalism\cite{ANO2,ANO2a,ANO2b}) and through the connection between self-dual theory and two-dimensional harmonic oscillator\cite{bazeia,CW}. In order to become the present work self-contained, it was organized as follow. In section 2, we will set forth the general Bopp's shift matrix method. In section 3, we will present the open string attached to a D3-brane in a presence of a magnetic field. In section 4, the two-dimensional NC harmonic oscillator will be obtained with the application of the general Bopp's shift matrix method in the usual commutative two-dimensional harmonic oscillator and, after that, the transformation proposed in \cite{bazeia,CW} will be applied in the NC version of the harmonic oscillator, which it allows us to obtain the well know result presented in section 3. At the last section, the conclusion and some discussion will be presented.

\section{The general Bopp's shift matrix method}
\label{sec:1}

Let a Lagrangian $L(q_i,\dot{q}_i)$ and a Hamiltonian $H(p_i,q_i)$ be related to each other in the usual way,
\be
L(q_i,\dot{q}_i)=p_i\,\dot{q}_i-H(p_i,q_i),~ \texttt{ with i = 1,2},\label{L_geral}
\ee
where $\,q_i\,$ are the configuration coordinates and $\,p_i\,$ are the canonical momenta conjugated to $q_i$. Let us arrange $q_i$ and $p_i$ in the phase space vector with the following ordering: $\xi_\alpha=(q_1,p_1,q_2,p_2)$. 
Once a total derivative added to the Lagrangian doesn't affect the equations of motion, we can rewrite the Lagrangian given in Eq.(\ref{L_geral}) in terms of the components of the phase space vector as
\be
L(\xi_\alpha,\dot\xi_\alpha)=\frac12\,\xi_\alpha\,{f}_{\alpha\beta}\,\dot\xi_\beta-H(\xi_\alpha),~~ \texttt{with~}\alpha = \texttt{1,2,3,4},\label{L_symp}
\ee
where ${f}_{\alpha\beta}$ are the elements of the symplectic matrix given by
\be
{f}=\left(
\begin{array}{cr}
-\varepsilon ~&~  0            \cr
 0           ~&~  -\varepsilon \cr 
\end{array}                \label{symcanmet}
\right),
\ee
where $\varepsilon$ is the 2$\times$2 antisymmetric block matrix whose elements are the two-dimensional Levi-Civita symbol with $\varepsilon_{12}=1$. 
Since in symplectic formalism the canonical momenta are included in the phase space coordinates $\xi_\alpha$, the Hamiltonian should be treated as a potential function. 
The Euler-Lagrange equations of motion can be obtained through the Lagrangian given in Eq.(\ref{L_symp}), namely
\be
\dot\xi_\alpha={f}^{-1}_{\alpha\beta}\,\frac{\partial{H}}{\partial\xi_\beta} ~,     \label{EqMovSymp2}
\ee
where ${f}^{-1}_{\alpha\beta}$ are the elements of the inverse of the symplectic matrix given by
\be
{f}^{-1}=\left(
\begin{array}{cc}
\varepsilon ~&~  0           \cr
 0          ~&~  \varepsilon \cr 
\end{array}                \label{symcanmetinv}
\right).
\ee
The components of Eq.(\ref{EqMovSymp2}) are the usual Hamilton equations of motion. The Poisson brackets, 
and its generalization to include constraints known as Dirac brackets are one of the possibles starting point to quantize a theory, 
since they introduce the quantum constant $\hbar$ through direct replacement of Poisson or Dirac brackets by the 
corresponding commutator. For two classical quantities $\,F(q_i,p_i)\,$ and $\,G(q_i,p_i)\,$ this replacement is represented by
\be
\{F,G\}\longrightarrow-i\hbar\big[\widehat{F},\widehat{G}\big],
\ee
where $\widehat{F}$ and $\widehat{G}$ are the corresponding quantum operators.
We can express the Poisson(Dirac) brackets of $\,F(\xi_\alpha)\,$ and $\,G(\xi_\alpha)\,$ through symplectic formalism with the help of the elements of the inverse of the symplectic matrix as
\be
\{F,G\}=\frac{\partial{F}}{\partial\xi_\alpha}\,{f}^{-1}_{\alpha\beta}\,\frac{\partial{G}}{\partial\xi_\beta}  ~. \label{PB_geral}
\ee
From Eq.(\ref{PB_geral}), we can calculate the Poisson(Dirac) brackets of the phase space coordinates directly as
\be
 \{\xi_\alpha,\xi_\beta\}={f}^{-1}_{\alpha\beta} ~. \label{PB_basic}
\ee
Putting back this term into the Eq.(\ref{PB_geral}), we get
\be
\{F,G\}=\frac{\partial{F}}{\partial\xi_\alpha}\,\{\xi_\alpha,\xi_\beta\}\,\frac{\partial{G}}{\partial\xi_\beta} ~. \label{PB_geral2}
\ee

One strategy that can be followed to incorporate non-commutativity in the subject is taking Eq.(\ref{PB_basic}) as a reference, realizing that the elements of ${f}^{-1}$ are sorted by  {\em direct} Poisson(Dirac) brackets (with $\alpha=\beta$) and {\em crossed} Poisson(Dirac) brackets (with $\alpha\neq\beta$) of the phase space coordinates, the latter are given by
\be
\{{q}_1,{q}_2\}=0~,~~~~\{{q}_1,{p}_2\}=0~,~~~~\{{p}_1,{q}_2\}=0~,~~~~\{{p}_1,{p}_2\}=0~. \label{PB_crossed}
\ee
We can then, with a convenient change of basis $~\xi_\alpha \rightarrow \wt\xi_\alpha~$, to drive the system description to an appropriate room which it allows to extend the formalism and add the non-commutative parameters into the system. So, in order to achieve this proposal, we just need  to keep unchanged the direct Poisson(Dirac) brackets and replace the zero elements corresponding to the crossed Poisson(Dirac) brackets by arbitrary non-commutative parameters $a_{ij}$ consistent with the problem to be investigated. In general way, the crossed Poisson(Dirac) brackets in the new basis will be given by
\be
\{\wt{q}_1,\wt{q}_2\}=a_{11}~,~~~~\{\wt{q}_1,\wt{p}_2\}=a_{12}~,~~~~\{\wt{p}_1,\wt{q}_2\}=a_{21}~,~~~~\{\wt{p}_1,\wt{p}_2\}=a_{22} \label{PB_news}
\ee
and they must be arranged in new matrix $\wt{f}^{-1}$ in way to respect its antisymmetry: 
\be
\wt{f}^{-1}=\left(
\begin{array}{cc}
\varepsilon &  \Theta \cr
-\Theta^T   &  \varepsilon
\end{array}\right) ~,  \label{inv_met}
\ee
where $\Theta$ is the building block whose elements are the arbitrary non-commutative parameters $a_{ij}$. 
The matrix $\wt{f}^{-1}$ will be the inverse of the symplectic matrix in non-commutative basis. 
After that, the Poisson(Dirac) brackets in non-commutative basis may be mapped via the following equation,
\be
\{\wt\xi_\alpha,\wt\xi_\beta\}=\wt{f}^{-1}_{\alpha\beta} ~,    \label{PB_news2}
\ee
which it reproduces the results given in Eq.(\ref{PB_news}).

The next step is to find the effect of the change of basis in the Lagrangian given in Eq.(\ref{L_symp}). We can verify that $\det(\wt{f}^{-1})=[1-\det(\Theta)]^2$. Then, provided that $\det(\Theta)\neq1$, we can proceed working with $\wt{f}$ and readily fill the kinetic term of the Lagrangian with non-commutative parameters replacing $~{f}_{\alpha\beta}~$ by $~\wt{f}_{\alpha\beta}$. Beyond that, we may consider the \textit{ansatz} to keep unchanged the shape of the Hamiltonian while it's expressed in terms of phase space coordinates $\,\wt\xi_\alpha\,$ of the non-commutative basis, that will effectively result in a change of the Hamiltonian. Considering the kinetic coefficient
\be
A_{\xi_\alpha}=\frac12\,\xi_\beta\,f_{\beta\alpha}~,       \label{kin_coeff_orig}
\ee
the change in the Lagrangian given in Eq.~(\ref{L_geral}) can be represented by
\be
L(\xi,\dot\xi)=A_{\xi_\alpha}\,\dot\xi_\alpha-H(\xi_\alpha) ~~\longrightarrow~~
\wt{L}(\wt\xi,\dot{\wt\xi})=\wt{A}_{\wt\xi_\alpha}\,\dot{\wt\xi}_\alpha-H(\wt\xi_\alpha) ~,  \label{L-LNC}
\ee
with $\wt{A}_{\wt\xi_\alpha}$ containing the elements of the symplectic matrix $\wt{f}_{\alpha\beta}$, now enhanced with non-commutative parameters,
\be
\wt{A}_{\wt\xi_\alpha}=\frac12\,\wt\xi_\beta\,\wt{f}_{\beta\alpha} ~.  \label{kin_coeff_NC}
\ee

The next step is to make feasible a transformation which it allows to return to the original commutative basis and then express $\wt{L}$ given in Eq.(\ref{L-LNC}) on this basis. As a starting point, we can realize that the relation between commutative basis and non-commutative basis can be readily obtained with the substitution $F=\wt\xi_\alpha$ and $G=\wt\xi_\beta$ in Eq.(\ref{PB_geral2}), it follows that,
\be
\{\wt\xi_\alpha,\wt\xi_\beta\}=\frac{\partial{\wt\xi_\alpha}}{\partial\xi_\kappa}\,\{\xi_\kappa,\xi_\lambda\}\,\frac{\partial\wt\xi_\beta}{\partial\xi_\lambda}  ~. \label{PBn-PBo}
\ee
Setting the matrix $R$ in order to perform the transformation of basis, original to non-commutative,
\be
\frac{\partial{\wt\xi_\alpha}}{\partial\xi_\beta}=R_{\alpha\beta}  ~.   \label{transxi_res}
\ee
Taking into account Eq.(\ref{PB_basic}) and (\ref{PB_news2}), we can re-express Eq.(\ref{PBn-PBo}) in matrix form as
\be
\wt{f}^{-1}=R\,f^{-1}\,R^T  ~.                                \label{f_f0inv}
\ee
We believe that the appendix \ref{appendixA} is the most appropriate place to perform the calculations that show us that relation given in Eq.(\ref{f_f0inv}) leads us to another relation that must be satisfied by $R$:
\be
R^{T}\wt{f}\,R=f                                          \label{f_f0}                                                   
\ee
and an expression for $R$ that satisfies these latter two relations:
\be
R=\sqrt{\wt{f}^{-1}f} ~.              \label{matrix_R}
\ee
It follows from Eq.(\ref{transxi_res}) for $R$ independent of $\xi_\alpha$ that
\be
\wt\xi_{\alpha}=R_{\alpha\beta}\,\xi_\beta  ~,   \label{linear}
\ee
which it is the general presentation of the Bopp's shifts. Considering this result, we can verify that the return to original basis causes the following change in  $\wt{A}_{\wt\xi_\alpha}$,
\be
\wt{A}_{\wt\xi_\alpha}=-\frac12\,\wt{f}_{\alpha\beta}\wt\xi_\beta=-\frac12\left(\wt{f}\sqrt{\wt{f}^{-1}f}\right)_{\alpha\beta}\xi_\beta~,
\ee
which can be written in compact shape by defining the antisymmetric $S$ matrix as
\be
S=-\frac12\,\wt{f}\sqrt{\wt{f}^{-1}f}  ~,  \label{matrix_S}
\ee
such that
\be
\wt{A}_{\wt\xi_\alpha}=S_{\alpha\beta}\,\xi_\beta  ~.  \label{Atilde}
\ee

Then, since the non-commutative parameters are incorporated into the system, we can go back to the commutative basis $~\wt\xi_\alpha \rightarrow \xi_\alpha$, now with non-commutative parameters participating on the dynamics of $\xi_\alpha$. As a consequence of the \textit{ansatz}, the non-commutative Hamiltonian expressed in terms of phase space coordinates of the commutative basis will change its shape, although it effectively remains the same, so that $\;H(\wt\xi)=\wt{H}(\xi)\,$. Considering the kinetic coefficients given in Eq.(\ref{kin_coeff_orig}) and Eq.(\ref{kin_coeff_NC}), a remarkable result, that can be obtained with the help of Eq.(\ref{f_f0}), is the invariance of kinetic term:
\be
\wt{A}_{\wt\xi_\alpha}\,\dot{\wt\xi_\alpha}=A_{\xi_\alpha}\,\dot\xi_\alpha  ~.  \label{Axi_res}
\ee 
Such that the Lagrangian with non-commutative parameters $\wt{L}$, given in Eq.(\ref{L-LNC}), can be expressed in commutative basis as
\be
\wt{L}(\xi,\dot\xi)=A_{\xi_\alpha}\,\dot\xi_\alpha-\wt{H}(\xi_\alpha) ~. \label{L_NC}
\ee
%%%
The equation of motion from Lagrangian given in previous equation is expressed by
\be
\dot\xi_\alpha=f^{-1}_{\alpha\beta}\,\frac{\partial{\wt{H}}}{\partial{\xi_\beta}}  ~, \label{EqMov_NC}
\ee
which it is identified as being the Hamilton's equations of motion that now carry the non-commutative parameters. 
The Lagrangian given in Eq.(\ref{L_NC}) can be directly expressed in the configuration space; then, disregarding a total derivative term, we can read
\be
\wt{L}(q_i,\dot{q}_i)=p_i\,\dot{q}_i-\wt{H}(p_i,q_i)~, \texttt{ with i = 1,2}, \label{L_NC_config}
\ee
which should be compared with our starting Lagrangian given in Eq.(\ref{L_geral}). So when we changed from commutative basis to non-commutative, so we can switch-on the non-commutativity and then return to the original commutative basis. We will find that all the NC-ingredients were transferred to the Hamiltonian, which they could be interpreted as being an external (unknown) potential, or a background field, or Lorentz symmetry breaking mechanism or even a mass generation mechanism\footnote{A work in progress.}.

\section{The bosonic string theory attached in a D3-brane}
\label{sec:2}

Consider a bosonic string theory in the presence of a D3-brane\cite{BS}, where the $B_{\mu\nu}$ in the 1,2 direction is a background antisymmetric tensor field and the coordinates of the brane are $x^0,\,x^1,\,x^2,\,x^3$; the remaining coordinates play no role. The bosonic open string sector with string ends attached to the D3-brane in the light cone frame, defined as

\be
\label{ST000}
x^{\pm}=x^0\pm x^3,
\ee
with the usual light cone choice of world sheet time,
\be
\label{ST010}
\tau = x^+.
\ee
\noindent The string action is

\be
\label{ST020}
{\cal{L}} = \frac 12 \int_{-L}^L d\tau \,d\sigma \left[\left(\frac{\partial x^i}{\partial \tau}\right)^2 - \left(\frac{\partial x^i}{\partial \sigma}\right)^2 + B_{ij}\left(\frac{\partial x^i}{\partial \tau}\right) \left(\frac{\partial x^j}{\partial \sigma}\right)\right].
\ee

\section{The origin of non-commutativity in the bosonic string theory}
\label{sec:3}
In order to propose a NC version of the string theory, we apply the NC symplectic induction method\cite{ANO2}\cite{ANO2a}\cite{ANO2b} or the general Boop's shifts matrix method\cite{MACN}, also presented in section \ref{sec:1}, into the two-dimensional commutative harmonic oscillator of unit mass and frequency, which it will be mapped into the bosonic open string theory with string ends attached to the D3-brane. The two-dimensional harmonic oscillator has its dynamic governed by the following Lagrangian,
\be
\label{ST030}
{\cal{L}} = \frac 12 \dot q_i^2 - \frac 12 q_i^2, \texttt{ with i = 1,2}.
\ee
The corresponding first-order Lagrangian is 
\be
\label{ST040}
{\cal{L}} = p_i\cdot q_i -{\cal{H}}(q_i,p_i),
\ee
with the symplectic potential given by
\be
\label{ST041}
{\cal{H}}(q_i,p_i)=\frac 12 p_i^2 + \frac 12 q_i^2,
\ee
where $p_i$ are the canonical momenta conjugated to $q_i$. The non-commutativity is introduced into the model changing the brackets among the phase-space variables are
\ba
\label{ST045}
\left\{\tilde {q}_i, \tilde {q}_j\right\} &=& 0,\nonumber\\
\left\{\tilde {q}_i, \tilde {p}_j\right\} &=& \delta_{ij},\\
\left\{\tilde {p}_i, \tilde {p}_j\right\} &=& \theta_{ij},\nonumber
\ea
%where $\theta_{ij}$ is the antisymmetric non-commutative parameter. We will employ the standard representation (the second representation given in Appendix \ref{appendixB}, %considering that previous Poisson(Dirac) brackets led to $\,\wt{q}_{12}=0$, $\,\Delta=0\,$ and $\,\wt{p}_{12}\,\varepsilon_{ij}=\theta_{ij}\,$ in Eq.(\ref{noncommut_stand})). Thus, %they are embraced by the inverse of the symplectic matrix in non-commutative basis, namely:
%%%%%%%%%%%%%
where $\theta_{ij}$ is the antisymmetric non-commutative parameter (We will employ the second representation given in Appendix \ref{appendixB}; by considering $\,\wt{q}_{12}=0$, $\,\Delta=0\,$ and $\,\wt{p}_{12}\,\varepsilon_{ij}=\theta_{ij}\,$ in the Eqs.(\ref{CPB_SR_qq} -- \ref{CPB_SR_qp}), they will exactly reproduce the previous brackets). Thus, from Eq.(\ref{noncommut_stand}), these latter are embraced by the inverse of the symplectic matrix in non-commutative basis, namely:
%%%%%%%%%%%%%
\be
\label{ST050}
{\wt{f}}^{-1}=\left(
\begin{array}{cc}
 0 ~&~    \delta_{ij}  \cr
 -\delta_{ij}    ~&~   \theta_{ij}\cr
\end{array}
\right).
\ee
The symplectic matrix in non-commutative basis is given by
\be
\label{ST060}
{\wt{f}}=\left(
\begin{array}{cc}
 \theta_{ij}    ~&~   - \delta_{ij}  \cr
 \delta_{ij}    ~&~  0 \cr
\end{array}
\right).
\ee
In contrast with symplectic matrix in commutative basis which is given by
\be
\label{ST061}
{f}=\left(
\begin{array}{cc}
 0              ~&~   - \delta_{ij}  \cr
 \delta_{ij}    ~&~  0 \cr
\end{array}
\right).
\ee
The non-commutative transformation matrices $~R=\sqrt{\wt{f}^{-1}\,f}~$ and $~S=-\frac12\,\wt{f}\,R~$ are written as
\be
\label{ST062}
R=\left(
\begin{array}{cc}
 \delta_{ij}             ~&~   0  \cr
 \frac12\,\theta_{ij}    ~&~  \delta_{ij} \cr
\end{array}
\right)~,~~~~~~
S=\left(
\begin{array}{cc}
 -\frac14\,\theta_{ij}    ~&~   \frac12\,\delta_{ij}  \cr
 -\frac12\,\delta_{ij}    ~&~  0 \cr
\end{array}
\right).
\ee
The transformation of phase space coordinates, from the old $~\xi_\beta=(q_j, p_j)$ to the new $\tilde\xi_\alpha=(\tilde{q}_i, \tilde{p}_i)$, and the transformation from the old phase space coordinates to the new kinetic coefficients $\wt{A}_{\tilde\xi_\alpha}=(\wt{A}_{\tilde{q}_i},  \wt{A}_{\tilde{p}_i})~$ are performed by matrix $R$ and the antisymmetric matrix $S$ as
\be
\label{transformations}
\wt\xi_\alpha=R_{\alpha\beta}\,\xi_\beta~~~~,~~~~
\wt{A}_{\wt\xi_\alpha}=S_{\alpha\beta}\,\xi_\beta~.  
\ee
It follows that
\ba
\tilde{q}_{i}&=&{q}_{i}~, \label{q-trans}\\
\tilde{p}_{i}&=&{p}_{i} +\frac12\,\theta_{ij}\,{q}_{j}~; \label{p-trans}\\
&& \nonumber \\
\wt{A}_{\tilde{q_i}}&=&\frac12\,{p}_{i}-\frac14\,\theta_{ij}\,q_j~, \\
\wt{A}_{\tilde{p_i}}&=&-\frac12\,{q}_{i}~.
\ea 
We also can make the usual check:
\be
\wt{A}_{\tilde {\xi}_\alpha}\cdot \dot {\tilde {\xi}}_\alpha=\frac12\,p_i\,\dot{q}_i-\frac12\,q_i\,\dot{p}_i~.
\ee
In agreement with the method presented in section \ref{sec:1} and Refs.\cite{ANO2}\cite{MACN}, the NC first-order Lagrangian is
\be
\label{ST065}
\tilde{\cal{L}} = \wt{A}_{\tilde {\xi}_\alpha}\cdot \dot {\tilde {\xi}}_\alpha - \cal{H}(\tilde {\xi}_\alpha),
\ee
and writing off the total time derivative, the NC first-order Lagrangian, Eq.(\ref{ST065}), is rewritten as
\be
\label{ST066}
\wt{\cal{L}}(q_i,\dot{q}_i)=p_i\,\dot{q}_i-\wt{\cal{H}}(q_i,p_i)~.
\ee
where $\wt{\cal{H}}(q_i,p_i)={\cal{H}}(\tilde{q}_i,\tilde{p}_i)$. As the symplectic potential ${\cal{H}}(\tilde {q}_i,\tilde {p}_i)$ is the NC version of the one given in Eq.(\ref{ST041}), namely,
\be
\label{ST067}
{\cal{H}}(\tilde {q}_i,\tilde {p}_i)=\frac 12 \tilde {p}_i^2 + \frac 12 \tilde {q}_i^2,
\ee
The Hamiltonian density obtained from Eq.(\ref{ST067}), Eq.(\ref{q-trans}) and Eq.(\ref{p-trans}) results in
\be
\wt{\cal{H}}(q_i,p_i)=\frac12\,p_i^2+\frac12\,q_i^2+\frac12\,p_i\,\theta_{ij}\,q_j+\frac18\,\theta_{ij}\,\theta_{ik}\,q_j\,q_k~.
\ee
From the equation-of-motion, $\dot{q}_i=\frac{\partial\wt{\cal{H}}}{\partial{p}_{i} }$, we obtain that
\be
p_i=\dot{q}_i-\frac12\,\theta_{ij}\,q_j.
\ee
What helps us to eliminate $p_i$ from the right hand side of the Eq.(\ref{ST066}) in order to calculate the second-order Lagrangian as
\be
\label{ST070}
\wt{\cal{L}}(q_i,\dot{q}_i)=\frac12\,\dot{q}_i^2-\frac12\,q_i^2-\frac{\theta_{ij}}{2}\,\dot{q}_i\,q_j~.
\ee
After that, the two-dimensional NC harmonic oscillator is mapped into a bosonic open string theory, since the analogy proposed in Refs.\cite{bazeia}\cite{CW} goes as follows:
\ba
\label{ST080}
\dot{q}_i &\rightarrow& \frac{\partial x^i}{\partial \tau},\nonumber\\
{q}_i &\rightarrow& \frac{\partial x^i}{\partial \sigma}.
\ea
Introducing these relations into the Lagrangian density (\ref{ST070}), we get

\be
\label{ST090}
\tilde{{\cal{L}}} = \frac 12 \left[\left(\frac{\partial x^i}{\partial \tau}\right)^2 - \left(\frac{\partial x^i}{\partial \sigma}\right)^2 - \theta_{ij}\left(\frac{\partial x^i}{\partial \tau}\right) \left(\frac{\partial x^j}{\partial \sigma}\right)\right].
\ee
Interpreting the NC parameter $\theta_{ij}$ as being the background antisymmetric tensor field $(-B_{ij})$, the bosonic open string sector with string ends attached to the D3-brane in a presence of magnetic fields is restored, whose its dynamic is governed by the Lagrangian given in Eq.(\ref{ST020}).

\section{Conclusion}
\label{sec:4}
We show that the non-commutativity presents into the bosonic open string attached to a D3-brane, in a presence of magnetic fields, can be obtained from a non-relativistic mechanical model, since the NC parameter could be interpreted as being the background antisymmetric tensor field $(-B_{ij})$. This was achieved introducing a NC parameter into the two-dimensional commutative harmonic oscillator and, from the NC version of harmonic oscillator, the coordinates were mapped to time and space field derivatives. From this approach, it was disclosed a connection between a mechanic non-relativistic system with a relativistic field theory; this opens new possibilities to investigate the features, not only present in the bosonic open string theory but also in other relativistic field theories from a simple scenario (mechanical model); for example, it is possible to investigate the consequences from the mapping of the two-dimensional commutative harmonic oscillator forced and/or damped into an alternative version of the bosonic open string attached to a D3-brane\footnote{A work in progress.}.

\appendix

\section{General properties of matrices \textit{R} and \textit{S}}
\label{appendixA}

The matrix $R$ satisfies the relation previously given in Eq.(\ref{f_f0inv}),
\be
R\,{f}^{-1}R^T=\wt{f}^{-1} \label{R_prop_1}.
\ee
In order to get the matrix $R$ from the latter equation, we should provide that $\det(R)\neq0$\footnote{Since $\det(f^{-1})=1$, it follows from Eq.(\ref{R_prop_1}) that $\det(\wt{f}^{-1})={[\det(R)]}^2$; thus we must request the same condition it was provided to obtain $\wt{f}$ from $\wt{f}^{-1}$.}, such that we can isolate $f^{-1}$ moving $R$ and $R^T$ to the other side. After this, calculating the inverse of the resulting expression, we obtain that
\be
R^{T}\wt{f}\,R=f \label{R_prop_2}.
\ee
The matrix product obtained by juxtaposing the Eq.(\ref{R_prop_1}) and Eq.(\ref{R_prop_2}) in a way that its right side is given by $\wt{f}^{-1}f$ is read as
\be
R\,{f}^{-1}R^T\,R^T\wt{f}\,R=\wt{f}^{-1}f.
\ee
The previous equation can be inverted to $R^TR^T$, and after a transposition it follows that
\be
RR=\wt{f}^{-1}\left(R^{-1}\wt{f}^{-1}fR^{-1}\right)^Tf.  \label{RR}
\ee
At this point, inputting the right side of the expression $\,\wt{f}^{-1}f=RR\,$ in the middle of the term between parenthesis of the Eq.(\ref{RR}), 
we get as output the same expression, $RR=\wt{f}^{-1}f$. It's like if you asked someone: Is that? and then he answered: Yes that is. It follows that
\be
R=\sqrt{\wt{f}^{-1}f}.  \label{R_matrix}
\ee 
The matrix $S$ previously defined in Eq.(\ref{matrix_S}) can be expressed with the help of matrix $R$ as
\be
S=-\frac12\,\wt{f}\,R.  \label{SfR}
\ee
To proof the antisymmetry of the matrix $S$, we can write its transpose as
\be
S^T=\frac12R^T\wt{f}.          \label{RTf}
\ee
On the other hand, from the Eq.(\ref{R_matrix}) squared, $\,RR=\wt{f}^{-1}f\,$, it follows that $\,\wt{f}\,R\,R=f$, which when it's faced with Eq.(\ref{R_prop_2}) results in
\be
R^T\wt{f}=\wt{f}\,R.               \label{fR}
\ee
Then, it follows that the matrix $S$ is antisymmetric:
\be
S^T=\frac12\,\wt{f}\,R=-S.       \label{S_antisim}
\ee

\section{Symplectic matrices representations}
\label{appendixB}

We will express the relevant matrices by building blocks, which was made possible by the choice for the ordering of the components 
of the phase space vector given early as $\xi_\alpha=(q_1,p_1;q_2,p_2)$. This is the called ``Paired representation''.
In order to embrace the direct (${i}={j}$) Poisson brackets in a compact shape, we will consider the following building block
\be
\varepsilon=
\left(
\begin{array}{rl}
\{\wt{q}_i~,~\wt{q}_i\} ~&~  \{\wt{q}_i~,~\wt{p}_i\}  \cr
\{\wt{p}_i~,~\wt{q}_i\} ~&~  \{\wt{p}_i~,~\wt{p}_i\}
\end{array}\right) ~=~
\left(
\begin{array}{rl}
0  ~&~   1  \cr
-1 ~&~   0
\end{array}\right) , ~~~~\texttt{ with i = 1,2} ,     \label{epsilon}
\ee
having the following properties, $\varepsilon^2=-\mathbb{1}$ \  and \ $\varepsilon^T=-\varepsilon$.
In order to embrace the four non-commutative parameters corresponding to the crossed (${i}\neq{j}$) Poisson brackets in a compact shape, we will consider the other building block
\be
\Theta=
\left(
\begin{array}{rl}
\{\wt{q}_1~,~\wt{q}_2\} ~&~  \{\wt{q}_1~,~\wt{p}_2\}  \cr
\{\wt{p}_1~,~\wt{q}_2\} ~&~  \{\wt{p}_1~,~\wt{p}_2\}
\end{array}\right)  ~, \label{Theta}
\ee
with its determinant represented by
\be
\det{\Theta}\equiv{b}^2 ~. \label{det}
\ee
Then assigning arbitrary values for the non-commutative parameters, we must be able to find that the following identities are satisfied:
\be
\Theta\,\varepsilon\,\Theta^T\varepsilon=\Theta^T\varepsilon\,\Theta\,\varepsilon=-b^2\,\I~.    \label{Theta_ident}
\ee
After that, we can build the relevant matrices modeled by blocks in this representation.
The symplectic matrix and its inverse in commutative basis take the following shape:
\be\arraycolsep=1.4pt\def\arraystretch{1.5}
{f}=\left(
\begin{array}{cr}
-\varepsilon  &  0 \cr
0             & -\varepsilon
\end{array}\right) 
~~~~,~~~~
{f}^{-1}=\left(
\begin{array}{rl}
\varepsilon  ~&~~  0 \cr
0            ~&~~ \varepsilon
\end{array}\right)  ~.     \label{origin_paired}
\ee
The symplectic matrix and its inverse in non-commutative basis take the following shape:
\be\arraycolsep=1.4pt\def\arraystretch{1.5}
\wt{f}=\frac{1}{1-{b}^2}\left(
\begin{array}{cc}
-\varepsilon                     ~~&~~ -\varepsilon\,\Theta\,\varepsilon \cr
\varepsilon\,\Theta^T\varepsilon ~~&~~ -\varepsilon
\end{array}\right)      
~~~~,~~~~
\wt{f}^{-1}=\left(
\begin{array}{cc}
\varepsilon &  \Theta \cr
-\Theta^T   & \varepsilon
\end{array}\right)  ~.    \label{noncommut_paired}
\ee
The matrices that perform transformations from commutative to non-commutative basis, $R$ defined in Eq.(\ref{matrix_R}) and $S$ defined in Eq.(\ref{matrix_S}) take the following shape:
\be\arraycolsep=1.4pt\def\arraystretch{1.5}
R=
\left(
\begin{array}{cc}
{r}\,\I                             ~~&~~ -\frac{1}{2{r}}\,\Theta\,\varepsilon \cr
\frac{1}{2{r}}\,\Theta^T\varepsilon ~~&~~ {r}\,\I
\end{array}\right)      
~~~~,~~~~
S=\frac{1}{2\,\sqrt{1-{b}^2}}
\left(
\begin{array}{cc}
{r}\,\varepsilon                                   ~~&~~ \frac{1}{2{r}}\,\varepsilon\,\Theta\,\varepsilon \cr
-\frac{1}{2{r}}\,\varepsilon\,\Theta^T\varepsilon  ~~&~~ {r}\,\varepsilon
\end{array}\right),           \label{R_S}
\ee
where
\be
{r}=\frac12\left(\sqrt{1+{b}}+\sqrt{1-{b}}\right)      \label{r_parameter}
\ee
and ${r}$ satisfies the following identities:
\ba
  {r}^2+\frac{{b}^2}{4\,{r}^2}&=&1                          \,, \label{r_ident1} \\	
	{r}-\frac{{b}^2}{2{r}}      &=&{r}\sqrt{1-{b}^2}            \,, \label{r_ident3} \\
  {r}-\frac{1}{2{r}}          &=&\frac{\sqrt{1-{b}^2}}{2{r}}  \,. \label{r_ident4} 
\ea
The consistency relation, $R^2=\wt{f}^{-1}f$, can be checked with the help of identities given in Eq.(\ref{Theta_ident}) and Eq.~(\ref{r_ident1}) and 
$S$ can be obtained with the help of the identities given in Eq.(\ref{Theta_ident}), Eq.(\ref{r_ident3}) and Eq.(\ref{r_ident4}).

Sometimes it's convenient to order the components of the phase space vector in the usual way as $\xi_\alpha=(q_1,q_2;p_1,p_2)$, that will be named ``Standard representation''.
In this representation, the direct (${i}={j}$) Poisson brackets  are embraced by the identity matrix $\I$; however, the crossed (${i}\neq{j}$) Poisson brackets corresponding to the non-commutative parameters must be arranged in a more complicated way:
\ba
\label{PB_SR}
\wt{q}_{12}&\equiv&\{\wt{q}_1~,~\wt{q}_2\}  ~, \label{CPB_SR_qq}\\
\wt{p}_{12}&\equiv&\{\wt{p}_1~,~\wt{p}_2\}  ~, \label{CPB_SR_pp}\\
\Delta&=&\left(
\begin{array}{cc}
0                           &  \{\wt{q}_1~,~\wt{p}_2\} \cr
-\{\wt{p}_1~,~\wt{q}_2\}     &             0
\end{array}\right) ~.                          \label{CPB_SR_qp}
\ea
After that, we can build the relevant matrices modeled by blocks in this representation. 
The symplectic matrix and its inverse in commutative basis take the following shape:
\be\arraycolsep=1.4pt\def\arraystretch{1.5}
{f}=\left(
\begin{array}{rr}
0   ~& -\I \cr
\I  ~&  0
\end{array}\right) 
~~~~,~~~
{f}^{-1}=\left(
\begin{array}{rr}
 0  ~~&~\I \cr
-\I ~~&~ 0
\end{array}\right)~. \label{origin_stand}
\ee
The symplectic matrix and its inverse in non-commutative basis take the following shape:
\be\arraycolsep=1.4pt\def\arraystretch{1.5}
\wt{f}=\frac{1}{1-{b}^2}\left(
\begin{array}{cc}
\wt{p}_{12}\,\varepsilon ~~&~~ -\I+\Delta^T \cr
\I-\Delta           ~~&~~  \wt{q}_{12}\,\varepsilon
\end{array}\right)        
~~~~,~~~~
\wt{f}^{-1}=\left(
\begin{array}{cc}
\wt{q}_{12}\,\varepsilon  ~~&~  \I+\Delta \cr
-\I-\Delta^T         ~~&~  \wt{p}_{12}\,\varepsilon
\end{array}\right)~.    \label{noncommut_stand}
\ee 
The matrices that perform transformations from commutative to non-commutative basis, $R$ defined in Eq.(\ref{matrix_R}) and $S$ defined in Eq.(\ref{matrix_S}) take the following shape:
\be\arraycolsep=1.4pt\def\arraystretch{1.5}
R=
\left(
\begin{array}{cc}
{r}\,\I+\frac{1}{2\,{r}}\,\Delta  ~~&~~ -\frac{\wt{q}_{12}}{2{r}}\,\varepsilon \cr
\frac{\wt{p}_{12}}{2{r}}\,\varepsilon  ~~&~~ {r}\,\I+\frac{1}{2\,{r}}\,\Delta^T
\end{array}\right)      
~~~~,~~~~
S=\frac{1}{2\,\sqrt{1-{b}^2}}
\left(
\begin{array}{cc}
-\frac{\wt{p}_{12}}{2\,{r}}\,\varepsilon ~~&~~ {r}\,\I-\frac{1}{2{r}}\,\Delta^T \cr
-{r}\,\I+\frac{1}{2{r}}\,\Delta     ~~&~~ -\frac{\wt{q}_{12}}{2\,{r}}\,\varepsilon
\end{array}\right),          \label{R_S_stand}
\ee
where, in this case, $b^2\,\equiv\,\wt{q}_{12}\cdot\wt{p}_{12}-\det(\Delta)\,$ and $\,{r}\,$ is given in Eq.(\ref{r_parameter}).

\end{document}